\documentclass[useAMS,usenatbib,usegraphicx]{mn2e}

\def\gtsim {>\kern-1.2em\lower1.1ex\hbox{$\sim$}~}   
\def\ltsim {<\kern-1.2em\lower1.1ex\hbox{$\sim$}~}   

\def \apj {ApJ}
\def \apjs {ApJS}
\def \aj  {AJ}
\def \aap {A\&A} 
\def \mnras {MNRAS}

\def \araa {ARA\&A}

\def \cite#1{\citealt{#1}}

\title[GRAPE-SPH Chemodynamical Simulation of Elliptical Galaxies II]{GRAPE-SPH Chemodynamical Simulation of Elliptical Galaxies II: Scaling Relations and the Fundamental Plane}
\author[Chiaki Kobayashi]{Chiaki Kobayashi$^{1}$\thanks{E-mail:
chiaki@MPA-Garching.MPG.DE}\\
$^{1}$Max-Planck-Institute for Astrophysics,
Karl-Schwarzschild-Str. 1, D-85741 Garching, Germany}

\begin{document}

\date{Accepted ----. Received 2004}

\pagerange{\pageref{firstpage}--\pageref{lastpage}} \pubyear{2003}

\maketitle

\label{firstpage}

\begin{abstract}
We simulate the formation and chemodynamical evolution of 128 elliptical galaxies using 
a GRAPE-SPH code that includes various physical processes that are associated with the formation of stellar 
systems: radiative cooling, star formation, feedback from Type II and Ia supernovae 
and stellar winds, and chemical enrichment.
We find that the star formation timescale controls when and where stars form in the contracting gas cloud, determines the effective radius at given mass, and is constrained by observation to be ten times longer than the local dynamical timescale.
We succeed in reproducing the observed global scaling relations under our CDM-based scenario, e.g., the Faber-Jackson relation, the Kormendy relation, and the fundamental plane.
An intrinsic scatter exists along the fundamental plane, and the origin of this scatter lies in differences in merging history.
Galaxies that undergo major merger events tend to have larger effective radii and fainter surface brightnesses, which result in
larger masses, smaller surface brightnesses, and larger mass-to-light ratios.
We can also reproduce the observed colour-magnitude and mass-metallicity relations, although the scatter is larger than observed.
The scatter arises because feedback is not very effective and star formation does not terminate completely in our simulations.
$\sim 25\%$ of accreted baryons are blown away in the simulations, independent of the assumed star formation timescale and initial mass function.
Most heavy elements end up locked into stars in the galaxy. The ejected metal fraction depends only on the star formation timescale, and is $\sim 2\%$ even to rapid star formation.
\end{abstract}


\section{INTRODUCTION}
\label{sec:intro}

The internal structure of galaxies,  
the spectrophotometric, chemical, and dynamical properties at 
various locations within a galaxy,  is determined by
the processes of galaxy formation and evolution.
Stars in a galaxy are fossils;
the star formation and chemical enrichment 
history of the galaxy are imprinted on their kinematics and chemical abundances.
The SAURON project with the William Herschel Telescope (\cite{bac01}; \cite{ems04}) is providing wide-field 
mapping of the kinematics and stellar populations of nearby galaxies, which will certainly give stringent 
constraints on galaxy formation and evolution.
Multiobject and integral field spectrographs are being developed also on 8-10m ground-based telescopes, which will provide the time evolution of such internal structure.
To infer the physical evolution processes from the observational data,
it is necessary to construct a realistic model, i.e., a three-dimensional chemodynamical model, 
and to compare the theoretical predictions with such observational data.

How elliptical galaxies form is a long-standing matter of debate.
Two competing scenarios for the formation of elliptical galaxies have been proposed; the monolithic collapse 
(e.g., \cite{lar74b}; \cite{ari87}), and the major merger (e.g., \cite{too77}; \cite{kau93}; \cite{bau96}; \cite{ste02}). 
In Kobayashi (2004, hereafter K04), we constructed
a self-consistent three-dimensional chemodynamical model of ellipticals, 
introducing
various physical processes associated with the formation of stellar systems;
radiative cooling, star formation, feedback of Type II and Ia supernovae (SNe II and SNe Ia), and stellar winds (SWs), and chemical enrichment.
We then argued that both formation processes should arise
to explain the observed variation in radial metallicity gradients.
The metallicity is enhanced in the central dense region, and the metallicity gradients are generated.
However, because merging events weaken the metallicity gradients, and because the secondary star burst induced by the mergers is not enough to regenerate them, galaxies that form monolithically have steeper gradients, while galaxies that undergo major mergers have shallower gradients.
Therefore no correlation is found between the mass and gradients, as in the observation (\cite{kob99}).

While the internal structure of elliptical galaxies is greatly affected by their merging histories, their global properties should be determined from their masses according to the scaling relations.
Differences in merging history may provide the scatter in these relations.
In this paper, we investigate whether our simulated galaxies follow the observed correlations; the Faber-Jackson relation, the Kormendy relation, the colour-magnitude relation, the mass-metallicity relation and the fundamental plane (FP).

The FP is a correlation of early-type galaxies with $2+n$ 
parameters (e.g., \cite{djo87}; \cite{dre87}) that reflect the internal structures, and is a clue to understand formation and evolution of early-type galaxies.
One possible interpretation of the FP, 
defined by central velocity dispersions $\sigma_0$, 
absolute effective radii r$_{\rm e}$, and 
surface brightnesses within an effective radius SB$_{\rm e}$
attributes it to a correlation of the mass-to-light 
ratio $M/L$ to the total luminosity, or equivalently, to the total galaxy 
mass (e.g., \cite{fab87}). 
The dependence of $M/L$ on the mass and luminosity stems from
the stellar metallicity and/or age (\cite{pah98}).
However, elliptical galaxies may not be homologous along the FP,
and there remains a dispersion that is not due to observational error.
This may be caused by metallicity, age, and/or dynamical disturbance.
The FP is observed up to $z\sim 0.5$ in clusters (\cite{kel97}, 2000), which is understood as
the evidence for passive evolution since $z \ltsim 1$. 
For field galaxies, a comparable correlation is observed up to $z\sim 1$ (\cite{tre01}; \cite{geb03}; \cite{ven03}; \cite{wel04}). 
The larger scatter than in clusters and the zero-point offset at higher redshifts can be interpreted as an age difference, although other possibilities such as dynamical disturbance have not yet been discussed.

The details of our GRAPE-SPH chemodynamical code were described in K04, and we briefly summarize them in \S \ref{sec:model}.
Following the discussion in K04, we vary the parameters controlling star formation and the initial mass function, and we show how these affect global properties and their correlations (\S \ref{sec:param}).
With the best parameter set, we show the scaling relations of simulated galaxies comparing with the observations.
We focus on the fundamental plane, and discuss the origin of the scatter in \S \ref{sec:correlation}.
\S 4 and \S 5 respectively contain the discussion and our conclusions.

\section{CHEMODYNAMICAL MODEL}
\label{sec:model}

The characteristics of our GRAPE-SPH code may be summarized as follows (see K04 for the detail).


i) The SPH method (\cite{mon92} for a review) is adopted, and the gravity is calculated in direct summation
using the special purpose computer GRAPE (\cite{sug90}).
The SPH formulation used in the code is almost the same as in \citet{nav93}.
The GRAPE-SPH code was originally written by \citet{nak03}, 
and is highly adaptive in space and time through individual smoothing lengths and individual timesteps.
The calculations were done with the GRAPE5 system in the National Astronomical Observatory of Japan and the GRAPE6 of the University of Tokyo.

ii) {\bf Radiative cooling} is computed using a metallicity-dependent cooling function.
For primordial gas ([Fe/H] $<-5$), we compute the cooling rates using the two-body processes of H and He, and free-free emission, as in \citet{kat96}.
For metal enriched gas ([Fe/H] $\ge -5$), we use a metallicity-dependent cooling function computed with the MAPPINGS III software (\cite{sut93}).
In this cooling function, the elemental abundance ratios are set to be constant for given [Fe/H] according to the relations found in the solar neighborhood. [O/Fe]$=0.5$ for Galactic halo stars for [Fe/H] $\le -1$, and solar values for [Fe/H] $\ge 0$. We interpolate between these values for $-1<$ [Fe/H] $<0$.

iii) Our {\bf star formation} criteria are the same as in \citet{kat92};
(1) converging flow; $(\nabla \cdot \mbox{\boldmath$v$})_i < 0$,
(2) rapid cooling; $t_{\rm cool} < t_{\rm dyn}$, 
and (3) Jeans unstable gas; $t_{\rm dyn} < t_{\rm sound}$.
The star formation timescale is proportional to the dynamical timescale ($t_{\rm sf} \equiv \frac{1}{c}t_{\rm dyn}$), where $c$ is our star formation timescale parameter.
We also adopt the probability criterion (\cite{kat92});
A random number between 0 and 1 is compared with the probability $P \equiv 1 - \exp\left[-\frac{\Delta t_{\rm sf}}{t_{\rm sf}}\right]$ in a time interval $\Delta t_{\rm sf} = 2$ Myr.

If a gas particle satisfies the above star formation criteria,
a fractional part of the mass of the gas particle turns into a star particle.
Since an individual star particle has a mass of $10^{5-7} M_\odot$,
it dose not represent a single star, but an association of many stars.
The mass of the stars associated with each star particle
is distributed according to an initial mass function (IMF).
We adopt a power-law IMF, $\phi(m) \propto m^{-x}$ (the slope $x=1.35$ gives the Salpeter IMF), which is invariant to time and metallicity.

In \S \ref{sec:param}, we discuss the dependence of our results on our free parameters; the star formation timescale $c$ and the slope of the IMF $x$.
In order to reproduce the observed radius-magnitude relation and mass-metallicity relation, we will choose $c=0.1$ and $x=1.35$ as a standard model.

\begin{table}
\caption{Number of simulated galaxies}
\label{tab:number}
\begin{tabular}{lcccc}
\hline
\footnotesize
 & run & ellipticals & dwarfs & cDs \\ 
\hline
high resolution ($N\sim60000$) & 13 & 14 &  4 & 0\\
low resolution ($N\sim10000$)  & 52 & 55 & 37 & 0\\
wider region ($N\sim60000$)    &  9 &  5 &  4 & 9\\
total                          & 74 & 74 & 45 & 9\\
\hline
\end{tabular}
\end{table}

\begin{table*}
\caption{Number of each types of galaxies with different star formation parameter $c$ and the slope of the IMF $x$.}
\label{tab:number2}
\begin{tabular}{lccccc|cccc}
\hline
\footnotesize
 & [E1] & [E2] & [E3] & [E4] & [E5] & [D1] & [D2]& [D3] & [D4]\\ 
\hline
A: c=1.0, x=1.10 &  5 & 18 & 19 & 25 & 11 & 20 & 13 &  9 & 4\\
B: c=0.1, x=1.10 & 10 & 17 & 15 & 33 & 16 & 20 & 18 & 13 & 2\\
C: c=0.1, x=1.35 &  8 & 15 & 12 & 31 & 20 & 15 & 16 & 10 & 2\\
\hline
\end{tabular}
\end{table*}

iv) For the {\bf feedback} of energy and heavy elements, we do not adopt the instantaneous recycling approximation.
Via SWs, SNe II, and SNe Ia, thermal energy and heavy elements are ejected from an evolved star particle as functions of time, and are distributed to all surrounding gas particles out to a constant radius of 1 kpc.
The ejected energy of each SW, SN II, and SN Ia are $\sim 0.2 \times 10^{51}$ erg depending on metallicity, $1.4 \times 10^{51}$ erg, and $1.3 \times 10^{51}$ erg.
We distribute this feedback energy in purely thermal form, although a fraction of it (given by a free parameter $f_{\rm kin}$) can be distributed in kinetic form as a velocity perturbation to the gas particles (see \cite{nav93}).
As shown in Fig.14 of K04, if we adopt $f_{\rm kin}=0.1$, the star formation efficiency is lower, the surface brightness of the final galaxy decreases at the centre, and metal-rich gas blows out, resulting in effective radii which are too large and metallicity gradients which are too shallow.

For the metals, the mass-dependent nucleosynthesis yields of SNe II and SNe Ia are taken from Nomoto et al. 1997ab.
The upper and lower mass limits of the IMF are $0.05$ and $120M_\odot$, respectively.
The progenitor mass ranges of SWs and SNe II are $8-120M_\odot$ and $8-50M_\odot$, respectively.
For SNe Ia, we adopt the single degenerate scenario with the metallicity effect (\cite{kob98}, 2000), where the progenitors are the Chandrasekhar WDs with an initial mass of $3-8 M_\odot$, and the lifetimes are determined from the lifetimes of the secondary stars with $0.9-1.5M_\odot$ and $1.8-2.6M_\odot$ for the red-giants and main-sequence systems, respectively.

v) The {\bf photometric evolution} of a star particle is identical to the evolution of a simple stellar population.
Spectra $f_\lambda$ are taken from \citet{kod97} as a function of age $t$ and metallicity $Z$.

vi) The {\bf initial condition} is a slowly rotating sphere with a CDM initial fluctuation generated by the COSMICS package (\cite{ber95}).
The cosmological parameters are set to be $H_0=50$ km s$^{-1}$ Mpc$^{-1}$, $\Omega_m=1.0$, $\Omega_\Lambda=0$, and $\sigma_8=1.0$.
The initial angular momentum is added as rigid rotation with the constant spin parameter $\lambda \sim 0.02$.
We set different resolutions and total mass;
the total mass of $\sim 10^{12} M_\odot$ (baryon fraction of $0.1$) with comoving radius of $\sim 1.5$ Mpc, and with $N \sim 10000$ and $60000$ (the half for gas and the rest for dark matter), which are the same as in K04.
The mass of a dark matter particle is $\sim 1.8 \times 10^8 M_\odot$ and $\sim 3.0 \times 10^7 M_\odot$, and the mass of 
a gas particle is $\sim 2.0 \times 10^7 M_\odot$ and $\sim 3.3 \times 10^6 M_\odot$, respectively.
In this paper, we add a new sample for cD galaxies with the total mass of $\sim 10^{13} M_\odot$ from wider initial conditions with radius $\sim 3$ Mpc and $N \sim 60000$. This mass resolution is similar to the lower resolution of K04 sample.


\section{RESULTS}
\subsection{EVOLUTION HISTORIES}
\label{sec:problem}

Simulating the chemodynamical evolution of 74 fields with different cosmological initial conditions, we obtain 128 galaxies at the present time (i.e., $t=13$ Gyr).
Depending on the CDM initial fluctuations, in some cases one galaxy forms in the centre of the field, in the others one galaxy and several subgalaxies.
We select galaxies having stellar masses in a 20 kpc sphere larger than $4.5 \times 10^7 M_\odot$.
Although many less-massive subgalaxies form, we discard them because our resolution 
is not enough to study them in detail.
We summarize the number of runs and the resulting galaxies in Table 1. 

Different galaxies undergo different evolution histories.
The difference is seeded in the initial conditions.
Galaxies form through the successive merging of subgalaxies with various masses. These vary between
a major merger at one extreme and a monolithic collapse of a slowly rotating gas cloud at the other.
We classify galaxies into the 5 classes according to their merging histories, defining a major merger with $f \gtsim 0.2$ at $z \ltsim 3$;
[E1] monolithic,
[E2] assembly,
[E3] minor merger,
[E4] major merger,
and
[E5] multiple major mergers (see K04 for the detail).
The numbers of galaxies of each class is slightly different from the adopted parameter set as summarized in Table 2, although the galaxy identification method is the same in K04.
The fractions of major merger galaxies among elliptical galaxies are 46\%, 54\%, and 59\% respectively for the model A, B, and C.
This is because the number of merging events increase at $1 \ltsim z \ltsim 3$
with slower star formation (smaller $c$) and weaker feedback (smaller $x$).

\begin{table*}
\caption{Total stellar mass, gas mass in the galaxy, gas mass outside the galaxy, the wind gas mass, and the total accreted baryon mass by the present in units of $M_\odot$.}
\label{tab:galaxy}
\begin{tabular}{lccccc|ccc}
\hline
\footnotesize
& $M_* (<r_{200})$ & $M_{\rm gas} (<r_{200})$ & $M_{\rm gas} (>2r_{200})$ & $M_{\rm wind}$ & $M_{\rm acc}$ & $M_{\rm wind}/M_{\rm acc}$ & $M_{\rm wind}/M_*$ & $M_{\rm Z,wind}/M_{Z,acc}$ \\
\hline
A: & $4.8 \times10^{10}$ & $3.4 \times10^{9}$ & $4.1 \times10^{10}$ & $1.85 \times10^{10}$ & $7.4 \times10^{10}$ & 0.25 & 0.39 & 0.023 \\
B: & $5.6 \times10^{10}$ & $3.7 \times10^{9}$ & $3.6 \times10^{10}$ & $1.91 \times10^{10}$ & $8.1 \times10^{10}$ & 0.24 & 0.34 & 0.008 \\
C: & $5.9 \times10^{10}$ & $1.9 \times10^{9}$ & $3.5 \times10^{10}$ & $1.92 \times10^{10}$ & $8.3 \times10^{10}$ & 0.23 & 0.33 & 0.010 \\
\hline
\end{tabular}
\end{table*}

As well as the observed dwarfs, the variation of the star formation histories is seen for our simulated dwarfs with $M_{\rm V,tot}\gtsim -19$ mag. We classify these into the 4 classes;
[D1] initial starburst,
[D2] continuous star formation,
[D3] continuous star formation with recent starburst,
and
[D4] recent starburst.
Observationally, the first class of galaxies would be dwarf ellipticals, the others are dwarf irregulars. 
The star formation rate (SFR) is truncated in [D1] because of the supernova feedback. 
The intermittent SFR is induced by the gas accretion and/or interaction 
with other galaxies.
In the following sections, we show the results both for giant and dwarf galaxies, but we should note that our numerical resolution in the dwarf galaxies is not enough; the particle number in a galaxy is small ($\sim 200$ in the worst case) and the gravitational softening ($1.0$ kpc for the low resolution) is comparable to the size of galaxies.

All simulated ellipticals form with an initial starburst at $z \gtsim 2$ with the typical timescale of $1-2$ Gyr.
The SFR decreases because the gas is exhausted in the galaxy.
The secondary starburst is induced by the accretion of gas clumps and/or the merging of gas-rich galaxies.
Not all merging events induce a secondary starburst; the fraction of merging events 
induce such a starburst is about $10\%$, depending on the gas mass of the secondary galaxy.
The initial starburst is always larger than the secondary one. 

As discussed in K04, such a truncating SFR is due to the artificial cut-off of mass accretion caused by the vacuum boundary of our initial conditions.
However, if we simulate wider region, the star formation continues longer, and thus colours tend to be too blue.
The mass accretion can be continued but the star formation should have stopped by some process by $z \sim 2$ in cluster ellipticals (e.g., \cite{kod97}; \cite{ell97}; \cite{kod98}; \cite{sta98}; \cite{bro00}) and $z \sim 1$ in field ellipticals (e.g., \cite{sil98}; \cite{sch99}; \cite{bri00}; \cite{dad00}; \cite{im02}).
An analogous truncation is required such as tidal stripping and effects of active galactic nuclei (AGN) (see \S \ref{sec:discussion}).
Eventually, although the dark matter halo may be affected, the stellar population does not change so much.

Contrary to the observational constraints, galactic winds are hard to generate in our simulations, and star formation does not completely stop by $z=0$ in most simulated elliptical galaxies.
At the centre of the present-day galaxy, the dynamical potential is so deep that the gas 
density is high. In such regions, super metal-rich stars ($Z\sim10Z_\odot$) keep on forming in the simulation.
In several dwarf galaxies with $M\ltsim$$10^9 M_\odot$, weak galactic winds can be seen even if $f_{\rm kin}=0$.
By the present epoch, roughly $25\%$ of the baryons are blown away by the input of thermal energy from supernovae, but most of the heavy elements remain locked into stars in the galaxy (see \S \ref{sec:param}).

\begin{figure}
\begin{center}
\includegraphics[width=8cm]{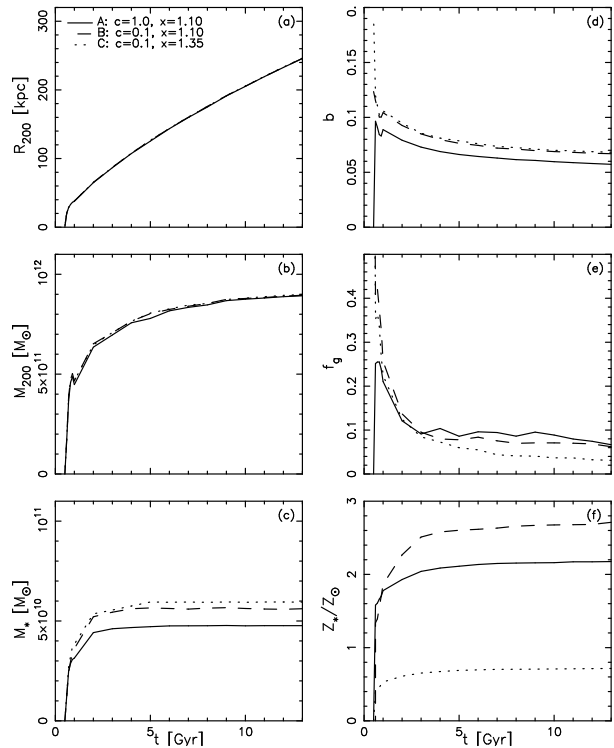}
\caption{\label{fig:param1}
Time evolution of global properties of a non-major merger galaxy;
(a) the radius $r_{200}$ with the mean density higher than $200\rho_{\rm c}$, (b) the total mass $M_{200}$ in $r_{200}$, (c) the total stellar mass $M_*$, (d) the baryon fraction $b \equiv (M_*+M_{\rm g})/M_{200}$, (e) the gas fraction $f_{\rm g} \equiv M_{\rm g}/(M_*+M_{\rm g})$, and (f) the mean stellar metallicity.
The solid, dashed, and dotted-lines are respectively for models A, B, and C.
}
\end{center}
\end{figure}

\begin{figure}
\begin{center}
\includegraphics[width=8cm]{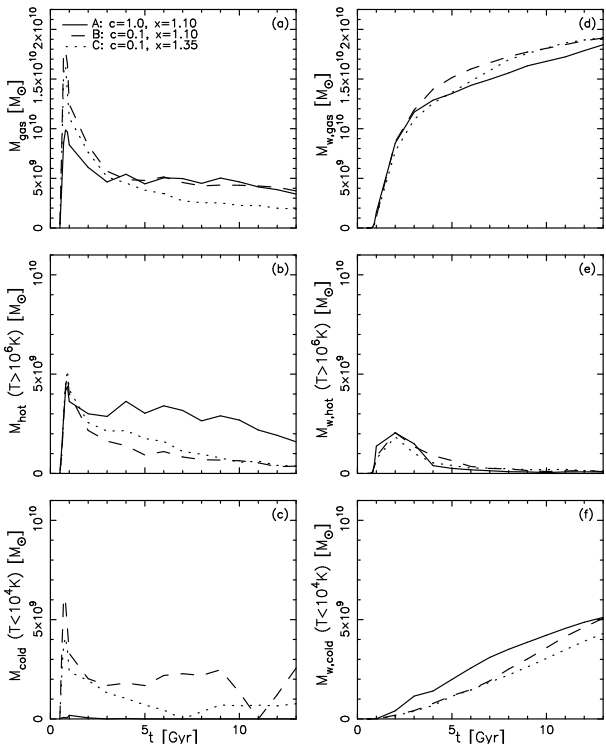}
\caption{\label{fig:param2}
Time time evolution of the gas mass in the galaxy (a-c) and in the wind (d-f) for all gas (a, d), hot gas with $T>10^6$ K (b, e), and cold gas with $T<10^4$ K (c, f). 
The wind gas is defined as the gas particles that have been inside $r<r_{200}$ and are outside $r>2r_{200}$ at present.
The solid, dashed, and dotted-lines are respectively for models A, B, and C.
}
\end{center}
\end{figure}

\subsection{PARAMETER DEPENDENCES}
\label{sec:param}

We show the parameter dependence of the global properties of a galaxy for varying star formation parameter $c$ and slope of the IMF $x$ in Figures \ref{fig:param1} and \ref{fig:param2}. 
The present values are summarized in Table 3.
We adopt identical initial condition for a non-major merger galaxy (ID 782389913, [E1] monolithic-like) with high resolution.
The model A (solid lines, $c=1.0, x=1.10$) is the same as in K04, where the feedback is stronger than the others because the star formation takes place at earlier epoch before the gas cloud collapses enough.
With the model B (dashed lines, $c=0.1, x=1.10$), the star formation takes place more slowly, and thus the stellar metallicity increases a lot.
With the model C (dotted lines, $c=0.1, x=1.35$), which is adopted in the following sections as a standard model, the stellar yield is reduced with smaller $x$, and thus the feedback is weaker than in the other models.
Here we define the galaxy as $r<r_{200}$, i.e., the region with the mean density higher than $200\rho_{\rm c}$ at each time.

Although the time evolution of the size and total mass (panels a and b) do not change at all, the stellar mass (panel c) varies with $c$ and $x$ by a factor of $20\%$ and $5\%$, respectively, in the sense that stronger feedback decreases the stellar mass.
The baryon fraction (panel d) is controlled by $c$.
Although the dark matter contraction is independent of these parameters, $c$ changes the peak epoch of the initial starburst ($0.3$ and $0.6$ Gyr respectively for $c=1.0$ and $0.1$), which determines when and where stars form in the contracting gas cloud.
On the other hand, the present gas fraction (panel e) does not depend on $c$, but on $x$.
This is because the return fraction is smaller with $x=1.35$ by a factor of $20-40\%$ than for $x=1.10$ depending on the lower mass limit $M_\ell$ of the IMF; 0.33 ($x=1.35$, $M_\ell=0.05M_\odot$, model C), 0.42 ($1.35$, $0.1M_\odot$), 0.45 ($1.10$, $0.05M_\odot$, model A), and 0.52 ($1.10$, $0.1M_\odot$, model B).
How much gas turns to stars finally depends both on $c$ and $x$ because some fraction of the gas is ejected from the galaxy.

Although the residual gas mass in the galaxy mainly depends on $x$, its temperature is also affected by $c$.
Figure \ref{fig:param2} shows the time evolution of the gas mass in the galaxy and in the wind for all gas (panels a, d), hot gas with $T>10^6$ K (panels b, e), and cold gas with $T<10^4$ K (panels c, f).
With the strongest feedback (model A, solid lines), the half of the residual gas is hot, and no cold gas exists.
With a longer timescale of star formation (model B, dashed lines), the residual gas can cool and the temperature decreases.
With model C (dotted lines), the gas cooling is not efficient because of lower metallicity.

In this simulation, although ten times larger amount of gas exists outside the galaxy ($r>2r_{200}$), some of them never accrete onto the galaxy.
We define the wind gas as the gas particles that have been inside $r<r_{200}$ and are outside $r>2r_{200}$ at present.
The wind gas mass is almost the half of the outside gas mass.
Although the temperature of the wind gas is different (panels e, f), the ejected wind mass by the present is independent of these parameters (panel d)
with an ejected wind fraction, the ratio of the wind mass to the accreted baryon mass $M_{\rm wind}/M_{\rm acc}$, of $\sim 25\%$.
The wind efficiency, the ratio of the wind mass to the total stellar mass $M_{\rm wind}/M_*$, depends on $c$, and is $\sim 0.4$ and $0.35$ for $c=1.0$ and $0.1$.
This means that it is easier for gas particles to be blown away when the gas collapse has not yet finished.

The biggest effect of $x$ is on the metallicity, and the mean stellar metallicity  of the whole of the galaxy (panel f) is $2.2$, $2.7$, and $0.7Z_\odot$ respectively for model A, B, and C.
The observational estimate using the metallicity gradients (\cite{kob99}) is $0.5-1$ solar.
Although $x=1.35$ is favored to meet this observational constraint under our star formation and feedback schemes, we cannot rule out other IMF.
In principle, $x=1.10$ gives stronger feedback, which can reduce the stellar metallicity. 
However, this feedback mechanism do not work well in our simulations and almost all the metals produced by supernovae are locked into stars.
Because the metallicity-dependent cooling function is included in our simulation, the metal enriched gas easily cools and loses the energy needed to escape.
In spite of the strong dependence of metallicity on $x$, 
the ejected metal fraction (the ratio of metal mass in the wind to the total produced metals $M_{\rm Z,wind}/M_{\rm Z,acc}$) depends only on $c$, and is $2\%$ and $1\%$ for $c=1.0$ and $0.1$, respectively.
With larger $c$, metals start to be lost before the galaxy collapses enough, and thus more metals can be ejected.
We should note, however, that these numbers are much smaller than expected from observations of X-ray gas in clusters.
These suggest that the ejected metal fraction should be larger than two-thirds (\cite{ren02}).

\begin{figure}
\begin{center}
\includegraphics[width=6cm]{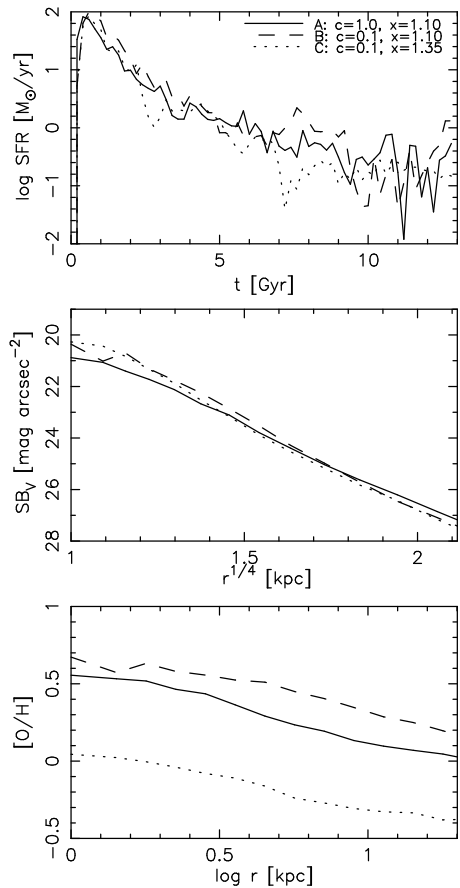}
\caption{\label{fig:param3}
The star formation rates derived from the ages of stars in the present-day galaxy (top panel), the surface brightness profile (middle panel), and the oxygen abundance gradients (bottom panel).
The solid, dashed, and dotted-lines are respectively for models A, B, and C.
}
\end{center}
\end{figure}

\begin{figure*}
\begin{center}
\includegraphics[width=16cm]{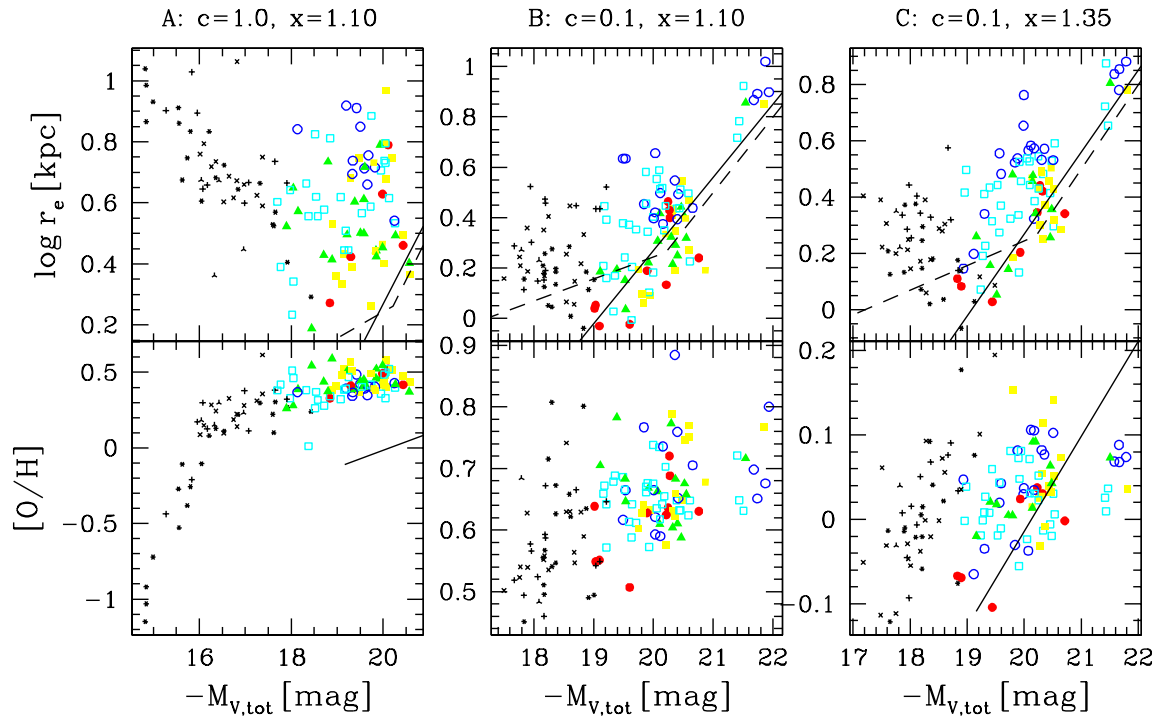}
\caption{\label{fig:allparam}
The relations of effective radius $r_{\rm e}$ (upper panels) and luminosity weighted metallicity (lower panels) to total V-magnitude for model A (left panels), B (middle panels), and C (right panels).
The solid and dashed lines show the observed relations of giant ellipticals (Pahre 1999) and dwarf ellipticals (Binggeli et al. 1984), respectively.
See the caption of Fig.\ref{fig:mass} for the symbols, which show the merging histories for elliptical galaxies (larger points) and star formation histories for dwarf galaxies (smaller points).
}
\end{center}
\end{figure*}

As discussed in K04, the galaxies of the K04 simulation are too extended, which causes an offset in the radius-magnitude relation.
This is because star formation takes place too early before the gas can accrete towards the centre. This problem can be solved by changing $c$.
Figure \ref{fig:param3} shows the star formation rates derived from the ages of stars in the present-day galaxy (top panel), the surface brightness profile (middle panel), and the oxygen abundance gradients (bottom panel).
The peak epoch of the initial star burst is delayed from $0.3$ Gyr with $c=1.0$ (solid lines) to $0.6$ Gyr with $c=0.1$ (dashed and dotted lines).
With $c=0.1$, the surface brightness increases at the centre, which results in  smaller $r_{\rm e}$ (4.3, 2.7, and 2.2 kpc respectively for models A, B, and C).
The metallicity gradient does not change.

We show scaling relations for our simulated galaxies in Figure \ref{fig:allparam}.
The effective radius $r_{\rm e}$ is derived by fitting the Vaucouleurs' law to the projection of $|Z|\le100$ kpc on the $X$-$Y$ plane (see K04 for the detail).
Figure \ref{fig:allparam} shows the relations of effective radius $r_{\rm e}$ (upper panels) and luminosity weighted metallicity (lower panels) to total V-magnitude.
The solid and dashed lines show the observed relations for giant ellipticals (\cite{pah99}) and dwarf ellipticals, respectively (Binggeli, Sandage \& Tarenghi 1984).
With model A (left panels), there is an off-set in the radius-magnitude relation because galaxies are too extended. The slope of the mass-metallicity relation is the same as observed, although the metallicity is too high.
With model B (middle panels), the radius becomes smaller by a factor of three, which agrees well with the observed radius-magnitude relation. However, the metallicity is still too high and the slope of the mass-metallicity relation is too shallow.
With model C (right panels), the radius-magnitude relation remains in good agreement with observations, and the metallicity decreases to meet the observations. Although the dispersion is larger, the simulated galaxies also follow the observed mass-metallicity relation.

\subsection{CORRELATIONS}
\label{sec:correlation}

\subsubsection{Scaling Relations}

\begin{figure*}
\begin{center}
\includegraphics[width=17cm]{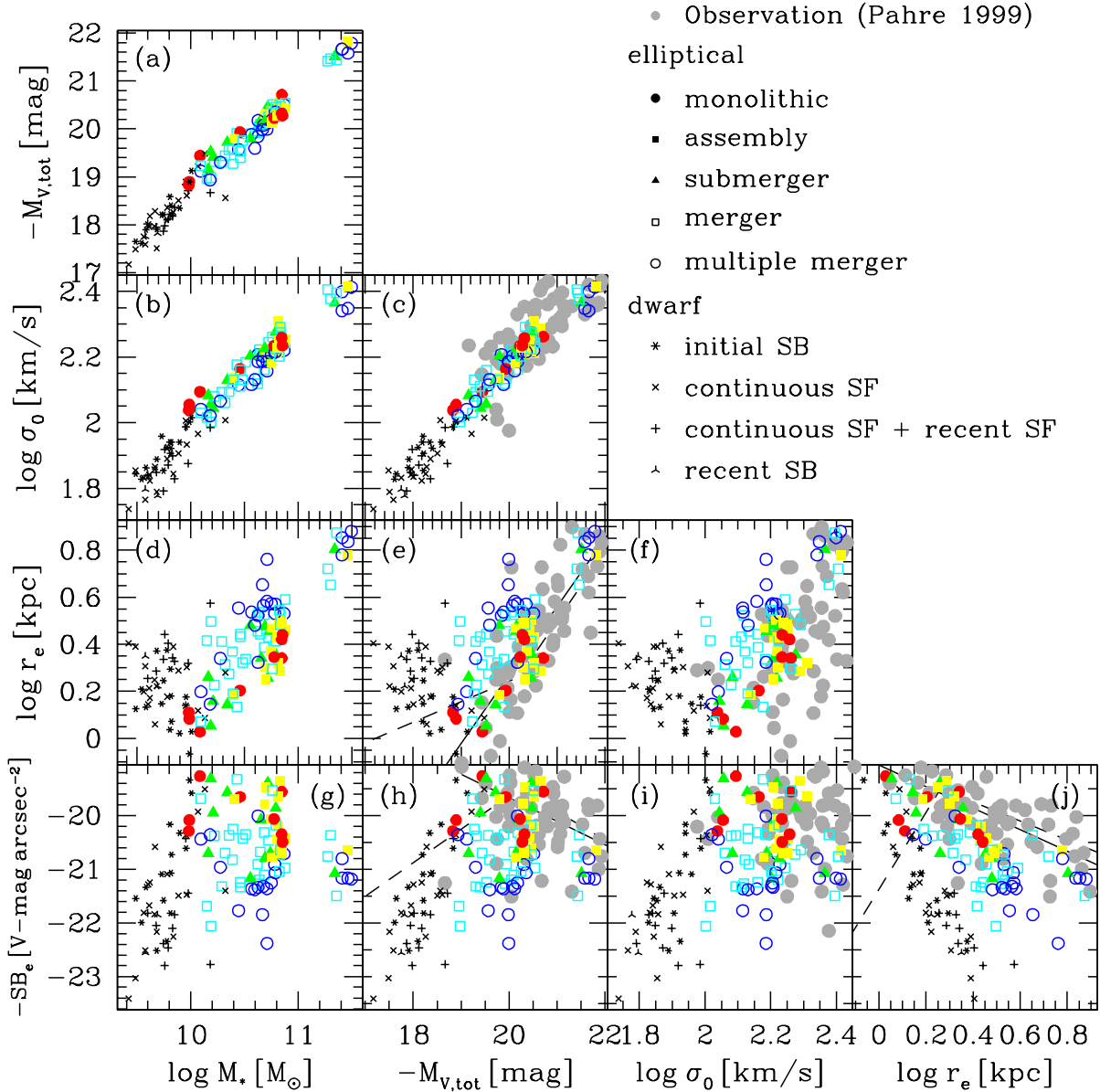}
\caption{\label{fig:mass}
The scaling relations among total stellar mass $M_*$ in $r_{200}$, total V-band luminosity $M_{\rm V,tot}$ derived from the de Vaucouleurs' law, central velocity dispersion $\sigma_0$ in 2 kpc, the effective radius $r_{\rm e}$, and mean surface brightness ${\rm SB}_{\rm e}$ within $r_{\rm e}$ in V-band.
All points are for the simulated galaxies, and massive and luminous galaxies are on the right side in all diagrams.
The symbols, showing the merging histories for elliptical galaxies and
star formation histories for dwarf galaxies, are: [E1] monolithic 
(filled circles), [E2] assembly (filled squares), [E3] minor merger (filled triangles), 
[E4] major merger (open squares), and [E5] multiple major merger (open circles);
[D1] initial starburst (asterisks), [D2] 
continuous star formation (crosses), [D3] continuous star formation with recent star 
burst (plus), and [D4] recent starburst (three-pointed stars).
The solid and dashed lines show the observed relations for early-type galaxies in the Coma cluster (Pahre 1999, also shown with the gray points) and for dwarf ellipticals in the Virgo cluster (Binggeli et al. 1984, B-V=0.9 is adopted), respectively.
}
\end{center}
\end{figure*}

For elliptical galaxies, it is well known that there are various correlations among physical scale parameters. 
The best known of these is the Faber-Jackson (1976) relation $L \propto \sigma^n$, where the slope is $n \sim 4$, but with a variation, depending on the sample definition (\cite{kor89}).
A correlation between the de Vaucouleurs' effective radius $r_{\rm e}$ and surface brightness SB$_{\rm e}$ was found by \citet{kor77}, where more luminous galaxies have larger $r_{\rm e}$ and fainter SB$_{\rm e}$.
However, the scatter of SB$_{\rm e}$ for a given $\sigma$ is quite large, and the SB$_{\rm e}-\sigma$ diagram corresponds to a face-on view of the fundamental plane (\cite{kor89}).
For dwarf galaxies, these relations are different;  the $M_B-r_{\rm e}$ relation has a shallower slope, and the $M_B-{\rm SB}_{\rm e}$ relation reverses with luminous dwarfs having brighter SB$_{\rm e}$ (\cite{bin84}).

Figure \ref{fig:mass} shows the scaling relations between total stellar mass $M_*$ in $r_{200}$, total V-band luminosity $M_{\rm V,tot}$ derived from the de Vaucouleurs' law, central velocity dispersion $\sigma_0$ in 2 kpc, effective radius $r_{\rm e}$, and mean surface brightness ${\rm SB}_{\rm e}$ within $r_{\rm e}$ in the V-band.
All points are for the simulated galaxies, and massive and luminous galaxies are on the right side in all diagrams.
The symbols show the merging histories for elliptical galaxies and star formation histories for 
dwarf galaxies as listed in the figure caption.

(panels a, b) {\it The total stellar mass} ($r<r_{200}$) correlates well with the total luminosity derived from the de Vaucouleurs' fit (panel a) and with the central velocity dispersion (panel b).
The correlations for the stellar mass measured in $r_{\rm e}$ have a scatter which is twice as large.

(panel c) {\it The Faber-Jackson relation} $L \propto \sigma^n$ shows a smaller dispersion than observed (gray points).
The slope is $n=3.8$ for this observation, but is steeper with $n=2.6$ for the simulation. 
This is because the simulated dwarf galaxies tend to have smaller $\sigma_0$ because of lack of resolution.
The slope for $-21<M_V<-19.5$ mag is consistent with $2.7$ both for the simulation and the observation.

(panels d, e, f) {\it The mass-effective radius relation}:
Massive/luminous ellipticals have larger effective radii.
The simulated galaxies follow the observed relation in the Coma cluster (\cite{pah99}, solid line). 
The dispersion is not so small as $\sim 0.7$ dex, but is comparable to the observation with $\sim 0.5$ dex (gray points).
In the simulation, the surface brightness of the central part is smeared by the gravitational softening, which causes an uncertainty in fitting the de Vaucouleurs' law.
For dwarf ellipticals, the observed relation has a shallower slope than that of giant ellipticals (\cite{bin84}, dashed line).
This tendency can be seen in the simulated dwarfs although the scatter is larger.

(panels g, h, i) {\it The mass-surface brightness relation}:
For giant ellipticals, massive/luminous galaxies tend to have smaller surface brightness, but this relation is reversed for dwarf ellipticals.
These tendency can be reproduced in the simulation, but the dispersion is very large.

(panel j) {\it The surface brightness-effective radius relation}:
The simulated giant ellipticals follow the observed relation where larger galaxies have lower surface brightnesses in $r_{\rm e}$. The scatter is almost the same as observed.
For dwarf galaxies, the observed relation is rectangular, and larger dwarfs have higher surface brightnesses.
The simulated dwarf galaxies populate the same side of the observed relation, but the direction of the relation is different from observation.
The effective radii of the simulated dwarfs tend to be too large, which is due to the lack of resolution.

The observed scaling relations are reproduced in the simulations.
The scatter exists even if the uncertainties of the simulations are taken account.
The origin of the scatter is clearly demonstrated by the symbols.
The galaxies that form monolithically (filled circles and squares) 
have smaller effective radii $r_{\rm e}$ and thus brighter surface brightness SB$_{\rm e}$, while the galaxies that undergo major mergers (open circles and squares) 
have larger $r_{\rm e}$ and thus fainter SB$_{\rm e}$.
This is because the merging events destroy the galaxy structures and make the radius larger and larger.
The dynamical information is not fully wiped out but is blurred by a merger, as is shown using the difference of the energies of particles before and after merging events in Figs.12 and 13 of K04.
Therefore, we conclude that the scatter of the scaling relations stems from differences in merging histories.

\subsubsection{Mass-Metallicity Relation}

\begin{figure}
\begin{center}
\includegraphics[width=9cm]{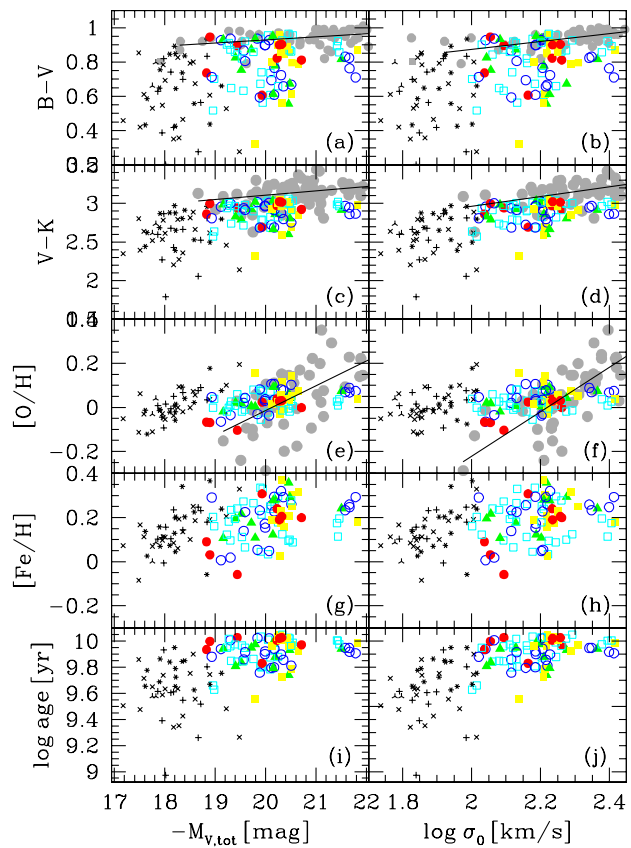}
\caption{\label{fig:metal}
Colours of B-V and V-K (a-d), oxygen and iron abundances (e-h), and ages (i-j),
which are measured in $r_{\rm e}$ weighted by V-luminosities,
versus total absolute V-magnitudes/central velocity dispersions.
The symbols are the same as in Fig.\ref{fig:mass}.
The solid line and gray points are for the observation taken from Bender et al. (1993) in (a-b) and Pahre (1999) in (c-f), respectively.
}
\end{center}
\end{figure}

The colour-magnitude relation of elliptical galaxies is
usually interpreted as luminous galaxies having higher stellar 
metallicities.
This is supported by the observation that a colour-magnitude relation with the same slope is found for high-redshift cluster ellipticals (e.g., \cite{kod97}).
Previous studies showed that the line index (Mg$_2$)$_0$ 
correlate with the velocity 
dispersion $\sigma_0$ at the galaxy centre (e.g., \cite{dav87}; Bender, Burstein \& Faber 1993)
and also with total absolute magnitude (e.g., \cite{bur88}).
The relation with $\sigma_0$ is tighter, but still shows some intrinsic scatter.
The same (Mg$_2$)$_0$-$\sigma_0$ relation is found both for cluster and field ellipticals (\cite{ber98}).
Therefore, the mass-metallicity relation of ellipticals is a common relation independent of environment and time, which contains important information on the star formation and feedback processes during the early stages of galaxy formation.

Figure \ref{fig:metal} shows the sequences of stellar populations against the galaxy mass. 
The mass tracers are the total absolute V-magnitude and the central velocity dispersion.
The characteristics of stellar populations are expressed by B-V and V-K colours, stellar abundances of oxygen [O/H] and iron [Fe/H], and stellar age, all of which are measured within $r_{\rm e}$ and weighted by V-luminosities.
Compared with observation, the mass-metallicity relations (panels e-h) are weak, with shallower slope and larger scatter in the simulations, although the average is consistent.
These are because the thermal feedback of supernovae is not enough to terminate star formation in the SPH simulations.
Especially, ejection of the metal enriched gas do not work well.
Although the steep slope of the observational relation requires not only more metal ejection in less-massive galaxies, but also more metal production in massive galaxies, such process cannot occur even if we change the IMF (\S \ref{sec:param}).
For the scatter, we find no significant dependence on merging history; it is mainly caused by the age differences because the luminosity weighted metallicity is affected by bright young populations formed in the secondary starbursts.
The relations for iron show larger scatter because iron is produced mainly by SNe Ia with longer lifetimes and affected by late star formation (see \S \ref{sec:discussion} for the abundance ratios).

Nonetheless, the majority of stars in the simulated giant galaxies are formed in an initial starburst, and the luminosity-weighted ages are as old as $7-10$ Gyr. No relation is found between age and galaxy mass (panels i-j).
For dwarf galaxies, the ages decrease to $3-8$ Gyr with a large scatter and a trend with mass can be seen, which is consistent with the observation using $H\gamma_\sigma$ index (\cite{yam04}).
The scatter stems from the differing star formation histories of dwarfs.
Among dwarf galaxies, dwarf ellipticals that formed by an initial star burst ([D1], asterisks) have larger ages up to $5-8$ Gyr.

Therefore, the simulated colour-magnitude relations (panels a-d) have shallower slope and a larger scatter than the observations.
At the reddest edge, we find the relation where massive ellipticals have red colours as observed.
Redder colours like V-K show smaller scatter because the origin of the scatter is the younger stars in the simulated galaxies.

\subsubsection{Mass-to-Light Ratios}

\begin{figure}
\begin{center}
\includegraphics[width=11cm]{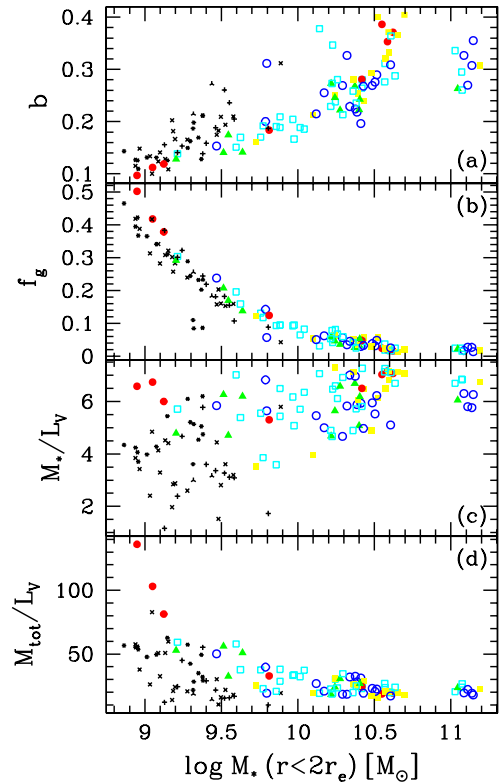}
\caption{\label{fig:mlb}
(a) Baryon fractions, (b) gas fractions, (c) stellar mass-to-light ratios, and (d) total mass-to-light ratios against stellar masses, in spherical regions with the radius of $2r_{\rm e}$. 
The symbols are the same as in Fig.\ref{fig:mass}.
}
\end{center}
\end{figure}

Figure \ref{fig:mlb} shows (a) baryon fractions, (b) gas fractions, (c) stellar mass-to-light ratios, and (d) total mass-to-light ratios against stellar masses.
All of them are measured in spherical regions with the radius of $2r_{\rm e}$, which includes $\sim 60\%$ of the total stellar mass.
The baryon fractions (panel a) are $\sim 0.35$ in massive galaxies, and decrease toward $0.1$ in dwarf galaxies.
In the galaxy, the baryon fraction increases toward the galaxy centre, but is $0.5$ at the most in the central  $2$ kpc.
Even at the galaxy centre, equal amounts of dark mass exist with the baryons in the simulated galaxies.
The gas fractions (panel b) are less than $5\%$ for the galaxies with $M_*>10^{10}M_\odot$, and increases towards $50\%$ in dwarf galaxies.
Therefore, the stellar mass to the total mass ratio well correlate with the galaxy mass. More stars form in more massive systems.

The stellar mass-to-light ratio (panel c) is derived in terms of the SSP model, which should be consistent with the observational estimates; 
for ellipticals, $M_*/L$ [$M_\odot/L_\odot$] $\sim 5-8$ and $6-9$ in the V-band and the B-band, respectively. 
A trend can be found that massive ellipticals have larger $M_*/L$, although the scatter is large.
For dwarfs, $M_*/L$ decreases to $2-6$ and $2-5$ in the V-band and the B-band, respectively, which is due to younger ages and lower metallicities than ellipticals.
However, in our simulation, the contribution of dark matter is quite large even at the galaxy centre. The total mass-to-light ratios (panel d) are as large as $20-40$ and $20-70$ respectively for the simulated ellipticals and dwarfs.
Observationally, the mass of dark matter can be estimated with the X-ray hot gas; the dark matter mass can be several times larger than the stellar mass for several galaxies even at the galaxy centre, which results in the total mass-to-light ratios of $\sim 20$ (\cite{mat98}).

\subsubsection{Fundamental Plane}

\begin{figure}
\begin{center}
\includegraphics[width=8cm]{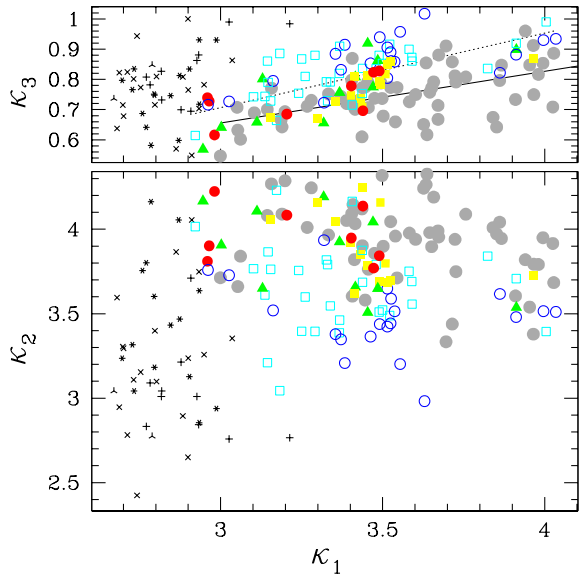}
\caption{\label{fig:fp}
The fundamental plane shown in $\kappa$-space for the V-band; the edge-on view (upper panel) and the face-on view (lower panel).
The symbols are the same as in Fig.\ref{fig:mass}.
The solid line and gray points shows the observed relation (Pahre 1999).
The dotted line show the rms fitting of the simulated ellipticals.
}
\end{center}
\end{figure}

Figure \ref{fig:fp} shows the V-band fundamental plane shown in the $\kappa$-space (\cite{ben92}). The parameters $\kappa_1$, $\kappa_2$, and $\kappa_3$ express masses, surface brightnesses, and mass-to-light ratios, respectively, and are defined as $\kappa_1\equiv(2\log\sigma_0+\log r_{\rm e})/\sqrt{2}$, $\kappa_2\equiv(2\log\sigma_0+2\log I_{\rm e}-\log r_{\rm e})/\sqrt{6}$, and $\kappa_3\equiv(2\log\sigma_0-\log I_{\rm e}-\log r_{\rm e})/\sqrt{3}$, where $\sigma_0$ is the central velocity dispersion and $I_{\rm e} \equiv 10^{-0.4({\rm SB}_{\rm e}-27)}$. 
The solid line shows the observed relation for the V-band (\cite{pah99}), and we reproduce the observed relations from the B-band to the near infrared.

The $\kappa_1$-$\kappa_2$ diagram (lower panel) is the face-on view of the fundamental plane. There is no correlation between masses and surface brightnesses, and the simulated galaxies cover the similar region to the observed giant and dwarf ellipticals (gray points, \cite{pah99}).
Dwarf galaxies populate the region with small masses and faint surface brightnesses.

The $\kappa_1$-$\kappa_3$ diagram (upper panel) is the edge-on view. 
The simulated galaxies follow the observed relation with a shallow slope (solid line, $\kappa_3=0.171\kappa_1+0.143$, \cite{pah99}), where more massive ellipticals have large ``mass-to-light ratios''.
Since the baryon fraction is as small as $0.5$ even in the galaxy centre, the ``mass-to-light ratios'' expressed by $\kappa_3$ is affacted by the dark matter content.
The rms fitting of the simulated ellipticals (dotted line), $\kappa_3=0.247\kappa_1-0.035$, is in agreement with the observation, although the slope is slightly steeper and the zero-point is larger by 0.1 dex than observed.
Dwarf galaxies lie above the relation, where the total mass-to-light ratios are larger than for giant ellipticals (Fig.\ref{fig:mlb}d).

\begin{figure}
\begin{center}
\includegraphics[width=8cm]{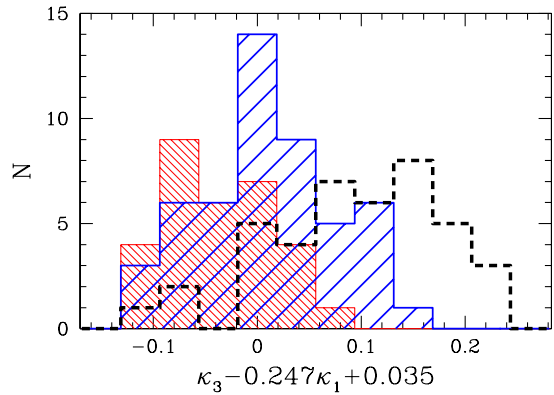}
\caption{\label{fig:hist}
Histograms of deviations from the fundamental plane.
The gray and hatched areas are for the simulated non-major merger ([E1]-[E3]) and major merger ([E4]-[E5]) galaxies, respectively.
The thick dashed line is for the simulated dwarf galaxies.
}
\end{center}
\end{figure}

An intrinsic scatter exists along the fundamental plane.
The origin of the scatter is clearly shown by the symbols;
merger galaxies (open circles and squares) have smaller $\kappa_2$ and larger $\kappa_3$ than non-merger ellipticals (filled circles and squares).
Figure \ref{fig:hist} shows the histograms of the deviations from the fundamental plane for the non-major merger ([E1]-[E3], gray area) and major merger galaxies ([E4]-[E5], hatched area).
These distributions are different, and major merger galaxies have larger $\kappa_3$ than non-major merger galaxies at given $\kappa_1$.
The thick dashed line is for the simulated dwarf galaxies, which have much larger $\kappa_3$.
Therefore, the origin of the scatter along the fundamental plane
is found to lie in differences in merging history.
As discussed for Fig. \ref{fig:mass}, 
the galaxies that undergo major mergers
tend to have larger $r_{\rm e}$ and fainter $I_{\rm e}$.
There is no significant change in $\sigma_0$ and total luminosity $L$.
From the definitions, these result in smaller $\kappa_2$ ($\propto\!L^2 \sigma_0^2  r_{\rm e}^{-5}$) and larger $\kappa_3$ ($\propto\!L^{-1} \sigma_0^2 r_{\rm e}$).

What is the origin of the fundamental plane?
In our simulation, the slope of the $\kappa_1$-$\kappa_3$ relation is originated from the combination of the metallicity, age, and the dark matter content.
i) {\it Metallicities}:
The slope mainly stems from the metallicity effect; metallicities are higher for massive galaxies (Fig.\ref{fig:metal}ef), which results in smaller total luminosities and larger stellar mass-to-light ratios (Fig.\ref{fig:mlb}c).
The scatter of the mass-metallicity relation is not small, and the scatter of the FP can be reduced by the other effects.
ii) {\it Ages}:
No relation is found between age and mass (Fig.\ref{fig:metal}ij), and 
galaxies are as old as $10$ Gyr coevally at $3.5 \ltsim \kappa_1 \ltsim 4$.
At $\kappa_1 \ltsim 3.5$, however, some low-mass metal-rich galaxies are younger and have $\kappa_3$ as small as the other metal-poor galaxies (e.g., the galaxies at $\kappa_1 = 3.25$ and $3.13$ on the dotted line).
iii) {\it Dark matter content}: 
Since the baryon fraction is larger for massive galaxies (Fig.\ref{fig:mlb}a), the total mass-to-light ratio is smaller for massive galaxies, which is the opposite of the FP.
However, the old and metal-rich galaxies at $\kappa_1 \sim 3.4-3.5$ have larger baryon fraction and less dark matter content, which results in smaller $\kappa_3$ than the massive galaxies at $\kappa_1 \sim 4$.

\section{DISCUSSION}
\label{sec:discussion}

The hydrodynamical simulation including star formation involves an uncertainty that is how to determine the star formation parameter $c$.
We show here that it can be constrained from the scaling relation of galaxies.
Because the star formation timescale controls when and where stars form in a contracting gas cloud in a dark matter halo,
the star formation timescale determines not only the ages of stars, but also the size of the galaxy.
To reproduce the observed radius-magnitude relation of elliptical galaxies, $c$ is constrained to be $0.1$ in our model, i.e., the local star formation timescale is ten times longer than the dynamical timescale.
The global star formation timescale is found to be $1-2$ Gyr, which is longer than  the $0.1$ Gyr that is commonly adopted in one-zone models (e.g., \cite{kod97}).
For spiral galaxies, the global star formation timescale should be several Gyr to meet various observations such as the metallicity distribution function of the Milky Way Galaxy.
Physical processes that change the star formation timescale might be the existence of rotation, the suppression of star formation due to the UV background radiation, and environmental effects in clusters.

We should note that mass accretion and star formation are truncated artificially by the initial conditions in our simulations.
However, in observed ellipticals, star formation should be truncated at $z \sim 2$ by some process, and an analogous truncation is required, as discussed in \S \ref{sec:problem}.
In our model, supernova feedback is not enough to stop the star formation, even if we increase the feedback energy with different IMF (\S \ref{sec:param}).
Hypernovae, which eject ten times larger energy than normal supernovae, may increase the supernova feedback.
Tidal stripping and ram-pressure stripping should affect, but it may be hard to explain the uniformity of cluster and field ellipticals.
The AGN feedback can suppress the star formation effectively.
If the relation between the black hole mass and the bulge luminosity (\cite{mag98}) suggests that the AGN activity increases following the merging of galaxies, the AGN feedback may become effective around this redshift.

To explain the lack of gas in present-day ellipticals, and to explain the heavy elements in the intracluster medium (\cite{cio91}), a galactic wind seems indispensable.
In our simulation, however, galactic winds do not occur in large galaxies, and thus star formation never terminates completely. This causes the large scatter of 0.3 dex in the B-V colour-magnitude relation (Fig. \ref{fig:metal}).
This problem arises from the SPH method and the feedback scheme.
Including kinetic feedback $f_{\rm kin}>0$ as in other simulations (\cite{kaw03}) does not seem to be a good solution.
With $f_{\rm kin}=0.1$, the surface brightness decreases at the centre of our ellipticals, and metal-rich gases blow out forming new stars at large radii. This results in effective radii which are too large and metallicity gradients which are too shallow to reproduce the observations.
Changing the IMF slope is not good solution either in our model.
$x=1.10$ gives stronger feedback, and the ejected wind mass does increase.
Unfortunately, the ejected metal fraction does not increase, and the overall stellar metallicity becomes too high.
The AGN feedback can not help solving problem because metals should be ejected during the star formation at very early epoch.
It takes time to generate super massive black hole by the merging of galaxies.
Because of the metallicity dependence of gas cooling, enriched gas is easily turn to stars.

Because of the existence of two distinct types of supernova explosion that produce different elements on different timescales, the abundance ratios of the stellar population can be used to put constraints on star formation histories.
SNe II, which are the core collapse-induced explosions
of short-lived massive stars ($\gtsim \, 8M_\odot$),
produce more O and Mg relative to Fe (i.e., [O/Fe] $>0$)
with a timescale of $10^{6-8}$ yr,
while SNe Ia, which are the thermonuclear explosions 
of accreting white dwarfs in close binaries,
produce mostly Fe and little O with a timescale of $0.5-20$ Gyr.
(The yields relative to solar value are almost the same between O and Mg both for SNe II and Ia.)
Observationally, it has often been claimed that Mg is
 overabundant in elliptical galaxies 
(\cite{wor92}; \cite{tho03}). 
Moreover, \citet{fis95}
showed that ellipticals with larger $\sigma_0$ tend to have larger
[Mg/Fe]$_0$.
However, in our simulation, luminosity weighted [O/Fe] spans $-0.3$ to $-0.1$ without any correlation with mass, although dwarf galaxies have larger [O/Fe] by 0.1 dex.
(In model A with stronger feedback and the flatter IMF with $x=-1.10$, [O/Fe] spans $-0.1$ to $0.1$.)
Star formation needs to be terminated by efficient feedback in the simulations.
We should note, however, that it may still be difficult to explain why [O/Fe] is larger in more massive galaxies.
The timescale and duration of star formation should be longer for massive galaxies
because the deep dynamical potential keeps gas cooling.
Some possibilities have been suggested (e.g., \cite{mat94});
i) the slope of the initial mass function may be different in massive ellipticals,
ii) the nucleosynthesis yields of SNe II may be different, namely, iron production may be different because of lower energies and/or larger fall-back,
iii) the binary frequency may be smaller and 
less SNe Ia may occur in massive ellipticals, and
iv) the metal enriched wind may cause selective mass loss, 
so that iron enriched gas can be ejected efficiently 
before it is consumed in forming the next generation of stars.

\section{CONCLUSIONS}

We study the formation and evolution of galaxies with a GRAPE-SPH chemodynamical model 
that includes various physical processes associated with the formation of stellar systems; radiative cooling, 
star formation, feedback from SNe II, SNe Ia, and SWs, and chemical enrichment.
We simulate 74 slowly-rotating spherical fields with CDM initial fluctuations (spin parameter $\lambda \sim 0.02$), and obtain 128 galaxies with stellar masses in the range $10^{9-12}M_\odot$ (74 ellipticals, 45 dwarfs, and 9 cD galaxies).
In our scenario, galaxies form through the successive merging of subgalaxies. 
The merging histories are various with differences seeded in the initial conditions.
In some cases, galaxies form through the assembly of gas rich small galaxies, and the process looks like a {\it monolithic collapse}. 
In other cases, the final galaxies form through a {\it major merger} of preexisting galaxies.
Major mergers are defined as those with mass ratio $f \gtsim 0.2$ occurring at $z \ltsim 3$.

Internal structure such as metallicity gradients is greatly affected by merging histories, while the global properties are determined from overall masses according to the scaling relations.
Assuming that the star formation timescale is ten times longer than the local dynamical timescale (i.e., $c=0.1$),
we succeed in reproducing the observed global scaling relations, e.g., the Faber-Jackson relation, the Kormendy relation, the colour-magnitude relation, the mass-metallicity relation and the fundamental plane.
The different relations for ellipticals and dwarfs could be reproduced, although simulated dwarfs have larger effective radii than observed because of the lack of resolution.
The luminosity-weighted ages of dwarfs span in wide range, $3-8$ Gyr, depending on their star formation histories, while ellipticals are as old as $7-10$ Gyr independent of their mass.

Adopting the Salpeter IMF ($x=1.35$), we could reproduce the mass-metallicity relations both for the central stellar metallicity and for the mean stellar metallicity of the whole of the galaxy.
However, the slope is shallower and the scatter is larger than observed, which are because the feedback is not so effective that most metals are locked into stars in the simulation.
The colour-magnitude relation also shows a larger scatter because the star formation does not terminated completely in the simulations.

An intrinsic scatter exists along the fundamental plane, and the origin of the scatter in the simulation lies in differences in merging history.
Galaxies that undergo major mergers tend to have larger effective radii and fainter surface brightnesses, which result in
larger $\kappa_1$ (expressing masses), smaller $\kappa_2$ (surface brightnesses), and larger $\kappa_3$ (mass-to-light ratios).

We examine the dependence of our results on the star formation parameter $c$ and the slope $x$ of the initial mass functions.
Although the time evolution of the size and total mass do not change at all, the stellar mass depends on these parameters in the sense that stronger feedback (i.e., larger $c$ and smaller $x$) decreases the stellar mass.
We found that $c$ controls when and where stars form in the contracting gas cloud, thus changing the baryon fraction
and determining the effective radius at given mass.
With the model to reproduce the observed mass-radius relation, the baryon fractions are $\sim 0.3$ and $\sim 0.1$, and the total mass-to-light ratios are $20-40$ and $20-70 M_\odot/L_\odot$ respectively for the simulated ellipticals and dwarfs.
The biggest effect of $x$ is on the metallicity, but it also changes the gas fraction and the residual gas mass in the galaxy.

On the other hand, the wind gas mass, which is defined as the gas particles that have been inside $r<r_{200}$ and are outside $r>2r_{200}$ at present, does not depend on these parameters, and $\sim 25\%$ of the accreted baryons can be blown away.
The wind efficiency, the ratio of the wind gas mass to the total stellar mass, depends on $c$, and is $0.4$ and $0.35$ for $c=1.0$ and $c=0.1$, respectively.
However, most heavy elements end up locked into stars in the galaxy.
The ejected metal fraction depends only on the star formation timescale, and is $\sim 2\%$ even if we take the quickest star formation rate (i.e., $c=1.0$) under our scheme.
To explain the metals detected in the intracluster medium in galaxy clusters, the feedback scheme should be improved so as that the enriched gas can blow away efficiently.
Changing the IMF do not help in solving this problem under our star formation and feedback schemes.

\section*{Acknowledgments}

This paper is a part of the Ph.D. thesis of C. Kobayashi in the Astronomy Department of the University of Tokyo.
I would like to thank the supervisor, K. Nomoto, N. Arimoto, and S.D.M. White  for detailed suggestions.
I am grateful to
N. Nakasato, J. Makino, T. Kodama, V. Springel, F. van den Bosch, and A. Renzini for fruitful discussions.
I also thank to the Japan Society for Promotion 
of Science for a financial support,
and to the National Observatory of Japan for the GRAPE system.

\bsp

\label{lastpage}

\end{document}